\documentstyle[aps]{revtex}

\def\rmdj {d\llap{\raise 1.22ex\hbox
  {\vrule height 0.09ex width 0.315em}\kern 0.04em}}
\def\rmDj {\rlap{\kern 0.05em\raise 0.76ex\hbox
  {\vrule height 0.10ex width 0.28em}}D}

\begin{document}
\title{Enhancement of Flux Pinning in Neutron Irradiated MgB$_{2}$ Superconductor}
\author{E. Babi\'{c}$^{\text{a}}$, \rmDj. Miljani\'{c}$^{\text{b}}$, K. Zadro$^{%
\text{a}}$, I. Ku\v{s}evi\'{c}$^{\text{a}}$, \v{Z}. Marohni\'{c}$^{\text{c}}$%
, \rmDj. Drobac$^{\text{c}}$, X.L. Wang$^{\text{d}}$ and S.X. Dou$^{\text{d}%
} $}
\address{$^{\text{a}}$Department of Physics, University of Zagreb, Bijeni\v{c}ka 32,\\
HR-1000 Zagreb, Croatia,\\
$^{\text{b}}$ Institute Ru\rmdj er Bo\v{s}kovi\'{c}, Bijeni\v{c}ka 54,\\
HR-10000 Zagreb, Croatia\\
$^{\text{c}}$ Institute of Physics, Bijeni\v{c}ka 46, HR-10000 Zagreb,\\
Croatia\\
$^{\text{d}}$ Centre for Superconducting and Electronic Materials, \\
University of Wollongong, Wollongong, NSW 2522, Australia}
\maketitle

\begin{abstract}
$m-H$ loops for virgin and neutron irradiated bulk and powder samples of MgB$%
_{\text{2}}$ were measured in the temperature range $5-30$ K in magnetic
field $B\leq 1$ T. The irradiation at thermal neutron fluences $9\cdot
10^{13}$ and $4.5\cdot 10^{14}$ cm$^{-2}$ caused very small enhancement of $%
m-H$ loops at lower temperatures ($T<20$ K), whereas the effect at high
temperatures was unclear due to difficulty in achieving exactly the same
measurement temperature prior and after irradiation. However, the
irradiation at $4.5\cdot 10^{15}$ cm$^{-2}$ produced clear enhancement of $%
m-H$ loops (hence $J_{c}$) at all investigated temperatures, which provides
the evidence for the enhancement of flux pinning in MgB$_{2}$ due to ion
tracks resulting from $\text{n}+^{10}$B reaction. The potential of this
technique for the enhancement of flux pinning in high temperature
superconductors is briefly discussed.
\end{abstract}

\bigskip 

The discovery of new superconductor\cite{1} MgB$_{\text{2}}$ aroused large
interest in the scientific community \cite{2}. Compared to high temperature
superconductors (HTS), MgB$_{\text{2}}$ has lower transition temperature, $%
T_{co}\cong 39$ K, but its simple composition, abundance of constituents and
probable absence of the problems associated with weak intergranular
couplings \cite{3,4} (inherent to HTS) make MgB$_{\text{2}}$ promising
material for the applications in the temperature range $20-30$ K which is
still well above $T_{c}$s for low temperature superconductors (LTS).
However, compared to LTS employed in the applications of superconductivity,
MgB$_{\text{2}}$ shows weaker flux pinning which manifests itself in sizably
lower critical current density\cite{5} $J_{c}$ at 4.2 K. Moreover, flux
pinning in MgB$_{\text{2}}$ decreases rapidly at elevated temperatures,
rendering quite low \cite{6} $J_{c}$ for $T\geq 20$ K. Since as of yet very
little effort has been put into optimizing $J_{c}$ of MgB$_{\text{2}}$,
there is ample space for the improvement of flux pinning in MgB$_{\text{2}}$.

In this respect, MgB$_{\text{2}}$ presents almost ideal testground for
testing our earlier proposal \cite{7} that flux pinning in boron doped HTS
and borocarbides \cite{8} can be enhanced via ion tracks associated with the
n + $^{\text{10}}$B reaction. For these purposes in the case of MgB$_{\text{2%
}}$ one can use $^{4}$He and $^{7}$Li ions from $^{10}$B(n,$\alpha $)$^{7}$%
Li reaction. Due to its large cross section at thermal energies ($\sigma
_{0} $ =3837 barn) and high abundance of $^{10}$B in natural boron
(19.78\%), one can reach high density of ion tracks in MgB$_{\text{2}}$
sample exposed to modest thermal neutron fluences. (This contrasts sharply
with U/n and Bi/p treatments employed for the increase of flux pinning in
HTS which require high fluences of thermal neutrons and high energy protons
respectively \cite{9}.) At thermal energies the reaction proceedes only via $%
^{7}$Li ground and first excited states with the $\alpha _{0}/\alpha _{1}$
branching of 6.7\%, i.e. the main contribution comes from the reaction
leading to the excited state. In this case the reaction products $^{4}$He
and $^{7}$Li nuclei have energies of 1.47 and 0.84 MeV respectively. The sum
of their ranges in MgB$_{\text{2}}$ with $\rho =2.6$ gcm$^{-3}$ is about 6 $%
\mu $m.

Large cross section $\sigma _{0}$ for $^{10}$B(n,$\alpha $)$^{7}$Li reaction
and high boron content in MgB$_{\text{2}}$ poses problems in irradiation of
bulk samples. The mean free path of thermal neutrons in MgB$_{\text{2}}$ is
about 0.2 mm. Obviously, in samples with thickness of $\sim 1$ mm the ion
tracks will be very unevenly distributed. In an isotropic neutron field more
than half of the ion tracks would be in the first 100 $\mu $m from the
surface. The problem is somewhat facilitated in thinner samples, but the
homogenous distribution of ion tracks in the samples is still unlikely.

The simplest way to monitor the enhancement of flux pinning (an increase in $%
J_{c}$) in type II superconductors is to measure the magnetization
hysteresis curves ($m-H$ loops). Since $J_{c}$ is proportional to
irreversible magnetization, any increase in the breadth of $m-H$ loop shows
the enhancement of flux pinning. This method can also be applied to powder
samples where more direct transport measurements of $J_{c}$ are not
possible. In what follows we present the preliminary results of
magnetization study performed on virgin and neutron irradiated bulk and
powder samples of MgB$_{\text{2}}$. Although these results were obtained for
rather thick samples ($d\geq 1$ mm) and quite low neutron fluences, they
indicate an enhancement of flux pinning following neutron irradiation both
in bulk and powder samples. Bulk MgB$_{\text{2}}$ samples were cut from
pellet prepared by conventional solid state reaction \cite{10}. SEM studies
revealed coarse grained structure with grain size of order\cite{10} of 200 $%
\mu $m. The measured density was $\sim 1$ gcm$^{-3}$, i.e. about 0.4 of the
ideal density of MgB$_{\text{2}}$. The sample for magnetization study had
dimensions 3.8$\times $1.5$\times $1 mm$^{3}$ and mass of 5.93 mg. Similar
sample (cut from adjoining part of the pellet) was used for ac
susceptibility study of superconducting transition temperatures prior and
after neutron irradiation. The powder sample was prepared by mixing small
amount of commercial (Alfa Aesar) MgB$_{\text{2}}$ powder (325 MESH size)
with about three times larger volume of epoxy in a plastic pill. After
setting, plastic container was removed and cylindrical sample with diameter
5.2 mm and length about 5 mm was split into two approximately
semicylindrical samples used for magnetization and ac susceptibility
measurements respectively. The $m-H$ loops were measured with commercial
vibrating sample magnetometer (VSM) in magnetic field up to 1 T at field
sweep rate 15 mT/s. Ac susceptibility was measured with highly sensitive ac
susceptometer \cite{11}. The employed temperature range was $5-40$ K. The
samples were irradiated at the roundabout of the TRIGA Mark II reactor of
the Jo\v{z}ef \v{S}tefan Institute in Ljubljana. At reactor power of 25 kW
the rotational irradiation facility has a flux of $1.45\cdot 10^{11} $ cm$%
^{-2}$s$^{-1}$ thermal ($E<0.5$ eV) and $0.22\cdot 10^{11}$ cm$^{-2}$s$^{-1}$
fast ($E>0.1$ MeV) neutrons. Initial thermal neutron fluence for all samples
was $9\cdot 10^{13}$ cm$^{-2}$. In later experiments, we employed fivefold
and fiftyfold larger thermal neutron fluences.

The ac susceptibility of bulk virgin sample (Fig. 1) showed superconducting
transition with diamagnetic onset at $T_{co}=38.2$ K and transition width $%
\Delta T_{c}(0.1-0.9)=0.46$ K in ac field amplitude of 1.5 A/m. In spite of
its porosity, sample showed excellent grain connectivity as manifested by
smooth, single step transitions at elevated ac fields ($\sim 10^{4}$ A/m).
The comparison of low field--low temperature diamagnetism in our sample with
that in niobium sample of approximately the same shape indicated Meissner
fraction $\geq 80\%$. The powder sample (Fig. 1) showed similar $T_{co}=38.2$
K (inset to Fig. 1), but most of transition was shifted to lower temperature
compared to that for bulk sample. At low field (1.5 A/m) this shift was 0.9
K at half transition, and the transition showed shallow tail bellow 35 K
which persisted down to 4.2 K (the transition in bulk sample was completed
at 35 K). These phenomena are consequence of a broad grain size distribution
(grain sizes $\leq 43$ $\mu $m) and impurity content in commercial MgB$_{%
\text{2}}$ powder. Both for powder and bulk samples, the superconducting
transitions after irradiations to $9\cdot 10^{13}$ and $4.5\cdot 10^{14}$ cm$%
^{-2}$ remained practically the same as those for virgin samples. In
particular, $T_{co}$ remained within 0.1 K of the initial value and the
breadths and magnitudes of the diamagnetic transitions practically did not
change. However, neutron fluence of $4.5\cdot 10^{15}$ cm$^{-2}$ caused
small shift of diamagnetic transition ($\approx 0.3$ K at half transition)
towards lower temperature for bulk sample (Fig. 1).

Fig. 2 shows magnetic moment vs. applied field loops for bulk sample at 5
and 10 K. The results for both virgin and irradiated (dashed line) sample
are shown. The field sweep rate was 15 mT/s. At elevated fields the
irradiation seems to produce marginal increase in the irreversible magnetic
moment and this increase is somewhat more pronounced at 10 K than at 5 K.
This result is consistent with rather low employed neutron fluence, uneven
distribution of ion tracks in our thick sample (most tracks concentrated
within thin surface layers) and much stronger (dominant) pinning effect of
intrinsic pinning centres which manifests itself in high critical current
densities of bulk MgB$_{\text{2}}$ samples at low temperatures \cite{5}.

However, the reappearance of the partial flux jumps (manifested as sharp
roughly zig-zag variation of magnetic moment at lower field in Fig. 2) in
irradiated sample at 10 K (note that virgin sample did not show flux jumps
at 10 K) seems to support the enhancement of flux pinning at 10 K after
neutron irradiation. These partial (pseudo) flux jumps have already been
reported for bulk MgB$_{\text{2}}$ samples similar to our one \cite{10} at
low temperatures ($T<10$ K). In contrast to normal flux jumps in LTS \cite
{12}, the above phenomenon is associated with violent flux entry in the
surface layers of porous sample. The phenomenon occurs only at low
temperatures where intragranular flux pinning greatly exceeds that in
intergranular links. Therefore, the reappearance of this phenomenon upon
irradiation seems to indicate an increase in the intragranular pinning
associated with ion tracks.

The $m-H$ loops for both virgin and neutron irradiated ($9\cdot 10^{13}$ cm$%
^{-2}$) powder sample were measured at temperatures 5, 10, 20 and 30 K,
respectively. For $T=5$ and 10 K the $m-H$ loops of irradiated powder were
identical to those of virgin one, whereas at 20 and 30 K $m-H$ loops of
irradiated sample showed small and sizable enhancement, respectively. No
effect of light irradiation on $m-H$ loops of MgB$_{2}$ powder for $T\leq 10$
K was plausible. It probably arised from combination of strong (dominant)
intrinsic flux pinning in MgB$_{2}$ powders (manifested by an order of
magnitude larger $J_{c}$s of MgB$_{2}$ powders compared to those of
corresponding bulk samples\cite{3,13}) and employed low neutron fluence
(hence low ion track density). However, large increase in the breadth of $%
m-H $ loops of irradiated powder at 30 K was unexpected. Although the
intrinsic flux pinning in MgB$_{2}$ grains decreases rapidly at elevated
temperatures \cite{6}, the pinning by ion tracks also becomes less efficient
due to corresponding increase of the coherence length of MgB$_{2}$\cite{5}
with temperature. Therefore, large effect of low neutron fluence on $m-H$
loops at 30 K seems unlikely.

Furthermore, rapid variations of $J_{c}$ with temperature for MgB$_{2}$
samples at elevated temperatures\cite{3,5,6,10,13} makes the comparison of
high temperature $m-H$ loops obtained in two experiments reliable only if
the sample temperatures were practically identical in two experiments. This
condition is difficult to fulfill with the temperature control technique
commonly used for VSMs. In VSM preheated helium gas flows around sample and
there is no thermometer at the sample holder. Therefore, the actual sample
temperature depends somewhat on the combination of the helium flow rate,
duration of the temperature stabilization and the size and properties (such
as shape, thermal capacity and conductivity, etc.) of the sample. Because of
this, we performed additional experiments in order to ascertain the role of
ion tracks in flux pinning in MgB$_{2}$. In the second experiment we
irradiated two powder and one bulk MgB$_{2}$ sample to five times larger
neutron fluence ($4.5\cdot 10^{14}$ cm$^{-2}$) than that used in first
experiment and measured their $m-H$ loops prior and after irradiation at
temperatures 10, 20 and 30 K. In order to minimize the eventual difference
in sample temperature in subsequent measurements we prepared plate-like
powder samples with thickness $d\approx 1$ mm and used longer time intervals
for the temperature stabilization ($t_{s}\geq 15$ min).

Fig. 3 shows $m-H$ loops both for virgin and irradiated ($4.5\cdot 10^{14}$
cm$^{-2}$) MgB$_{2}$ powder at temperatures 10, 20 and 30 K, respectively.
The breadth of $m-H$ loop after irradiation is slightly larger at 10 K,
shows almost no change at 20 K and appears somewhat smaller at 30 K. The
results for other powder sample were practically the same. Whereas the
results for 10 K and 20 K can be interpreted in terms of decreasing
efficiency of ion track pinning on increasing temperature, the decrease of $%
m $ (hence $J_{c}$) of irradiated sample at 30 K is at variance with the
fact that irradiation produced no measurable change in either transition
temperature or the shape of transition for MgB$_{2}$ powders. No change in
superconducting parameters of the samples is difficult to reconcile with the
suppression of their $J_{c}$s at 30 K. The most probable explanation of this
observation (Fig. 3) is that our control of the sample temperature, although
improved, is still insufficient for the reliable measurements of small
changes in $J_{c}$ at 30 K. The $m-H$ loops of bulk MgB$_{2}$ sample
irradiated to the same neutron fluence was at 20 K almost unchanged compared
to that for virgin one, whereas at 30 K it showed small enhancement. No
enhancement of $m-H$ loop at 20 K probably indicates that the enhancement at
30 K was at least partially due to a slightly lower measurement temperature
after irradiation. However, the $m-H$ loops of irradiated sample showed
strong flux jumps at 10 K (which were not observed in virgin sample) thus
clearly indicating an enhancement of flux pinning (larger $J_{c}$\cite{14})
after irradiation.

Taken together, these experiments show small enhancement of flux pinning at
lower temperatures ($T<20$ K) in irradiated MgB$_{2}$ samples (both bulk and
powder ones). However, for the employed (low) neutron fluences the effects
are to small to allow more quantitative assessment. Therefore, we performed
third experiment in which two powder and one bulk MgB$_{2}$ sample were
irradiated to neutron fluence $4.5\cdot 10^{15}$ cm$^{-2}$. After
irradiation the superconducting transition of bulk sample was shifted a
little towards lower temperature, the shift at half transition being about
0.3 K (Fig. 1). Since the powder samples showed an induced radioactivity,
probably caused by some impurity present in commercial MgB$_{2}$ powder
(95.5\% purity), the measurements on these samples were postponed.

Fig. 4 shows $m-H$ loops both for irradiated and virgin bulk MgB$_{2}$
sample at temperatures 10, 20, 25 and 30 K, respectively. The appearance of
flux jumps at 10 K and the increase in the breadths of $m-H$ loops for
irradiated sample at all other temperatures prove the enhancement of flux
pinning after irradiation. We also observe that the enhancement of $J_{c}$
for $T\geq 25$ K decreases with increasing temperature, which is partially
due to the decrease of $T_{c}$ upon irradiation (Fig. 1). The enhancement of 
$J_{c}$ due to ion track flux pinning at 20 K is about $2\%$ and $5\%$ at $%
\mu _{0}H=0$ and 0.6 T, respectively. Clearly, the enhancement of $J_{c}$
will become larger at higher applied fields. The observed rather modest
enhancement of $J_{c}$ in irradiated bulk MgB$_{2}$ sample is probably a
consequence of quite large coherence length in MgB$_{2}$, small masses and
energies of ions and the use of still rather modest neutron fluence. Since
in spite of porosity of our sample the decrease of $T_{c}$ is quite small,
the use of sizably larger neutron fluences seems feasible, which would
result in correspondingly larger enhancement of flux pinning. The other
advantage of flux pinning by ion tracks resulting from $\text{n}+^{10}$B
reaction is that $J_{c}$ is enhanced at all fields (including $H=0$, Figs. 3
and 4), whereas the other ion tracks invariably suppress $J_{c}$ at $H=0$
and the enhancement of $J_{c}$ occurs only at elevated fields\cite{9}.
Furthermore, the enhancement of flux pinning by ion tracks from $\text{n}%
+^{10}$B reaction in HTS (such as Bi--Sr--Ca--Cu--O compounds) should be
sizably larger due to considerably smaller coherence lengths and weaker
intrinsic flux pinning in these materials. Taken together, our results
confirm the flux pinning effect of ion tracks in MgB$_{2}$ and moreover
enable us to predict that the same technique may become powerful method for
the enhancement of flux pinning in HTS\cite{7}.

Acknowledgements

We acknowledge the expert help of Drs M. Ravnik, T. \v{Z}agar as well as of
the TRIGA reactor crew in neutron irradiation of bulk and powder samples of
MgB$_{\text{2}}$. We also thank Drs A.H. Li and S. Sontanian for help in
preparation of bulk MgB$_{\text{2}}$ sample.

\newpage
\begin{figure}[tbp]
\caption{Variations of real ($\chi '$) and imaginary ($\chi ''$) part of ac
susceptibility for commercial MgB$_{2}$ powder embedded in epoxy (dashed)
and bulk MgB$_{2}$ sample prior (full) and after irradiation at thermal
neutron fluence $4.5\cdot 10^{15}$ cm$^{-2}$ (dotted) with temperature for
alternating field amplitude $H_{m}=15.8$ A/m. The inset: enlarged view of
the onset of superconductivity for the same samples.}
\label{key}
\end{figure}

\begin{figure}[tbp]
\caption{Magnetization loops for bulk MgB$_{\text{2}}$ sample at 5 and 10 K.
Dashed line denotes results obtained after irradiation of sample to thermal
neutron fluence of $9\cdot 10^{13}$ cm$^{-2}$.}
\label{key}
\end{figure}

\begin{figure}[tbp]
\caption{Magnetization loops for MgB$_{\text{2}}$ powder embedded in epoxy at
10, 20 and 30 K, respectively. Dashed line is for the same sample exposed
to neutron fluence of $4.5\cdot 10^{14}$ cm$^{-2}$.}
\label{key}
\end{figure}

\begin{figure}[tbp]
\caption{Magnetization loops for bulk MgB$_{2}$ sample at 10, 20, 25 and 30 K,
respectively. Dashed line is for the same sample exposed to the neutron fluence
$4.5\cdot 10^{15}$ cm$^{-2}$.}
\label{key}
\end{figure}

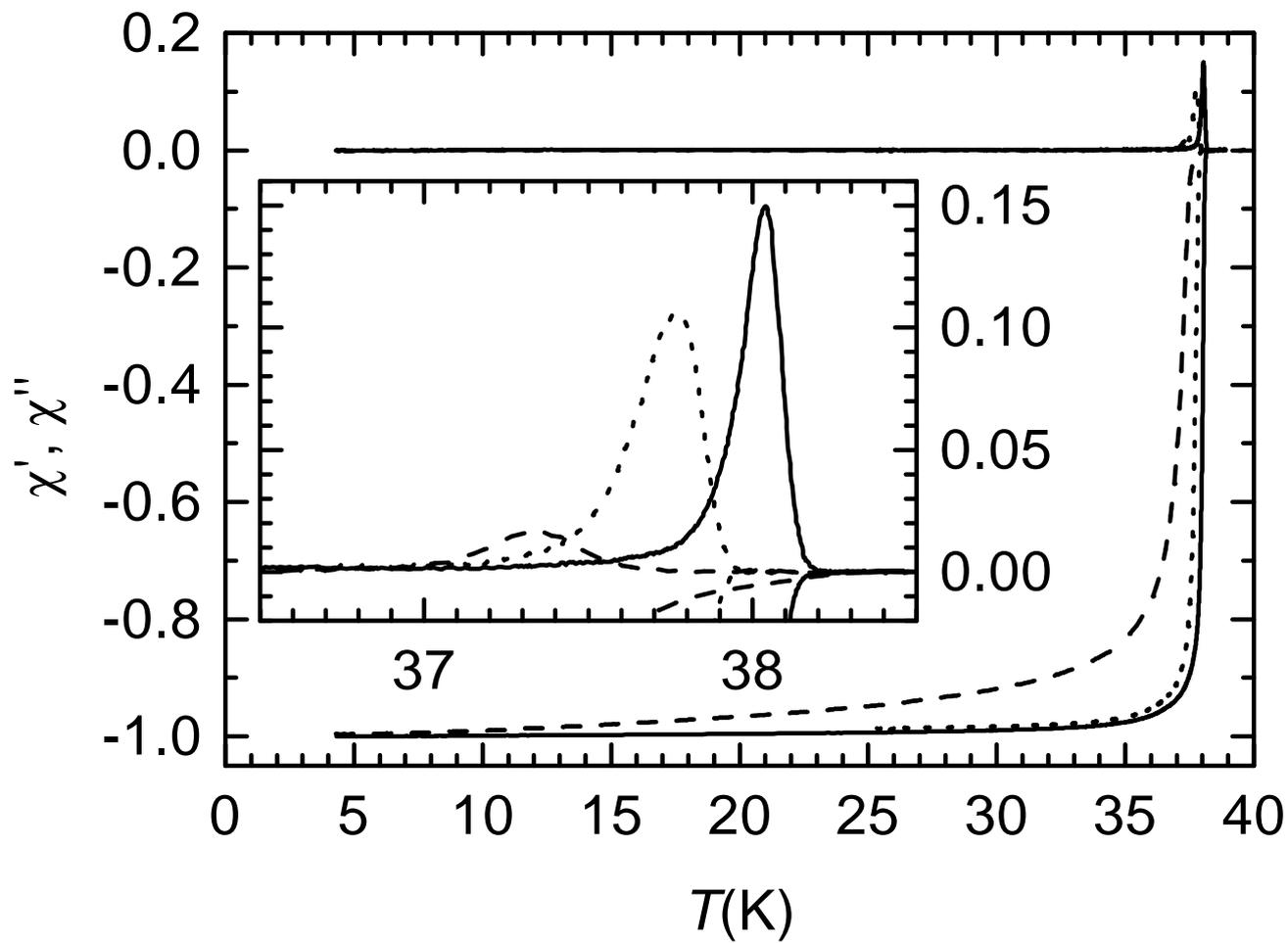
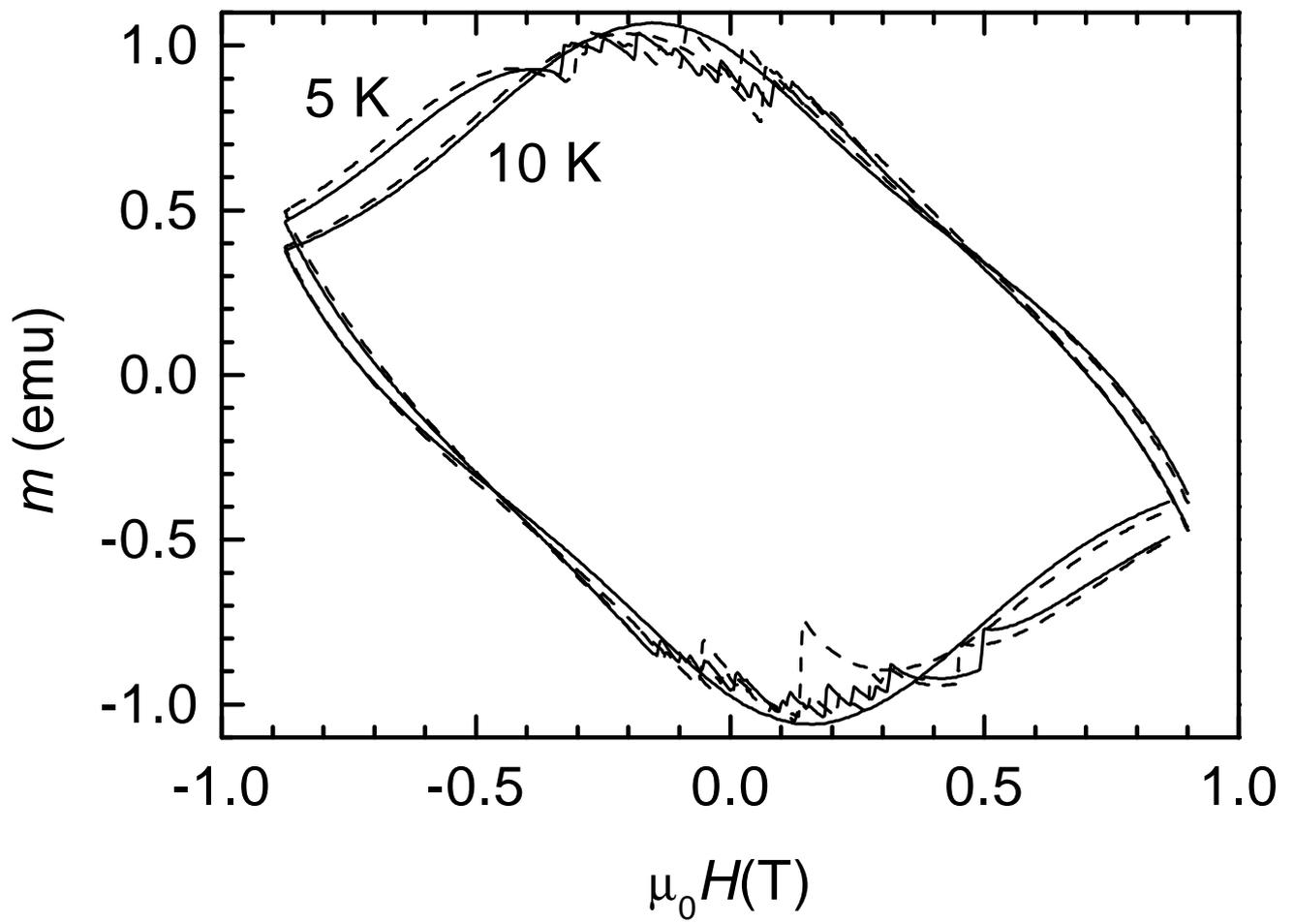
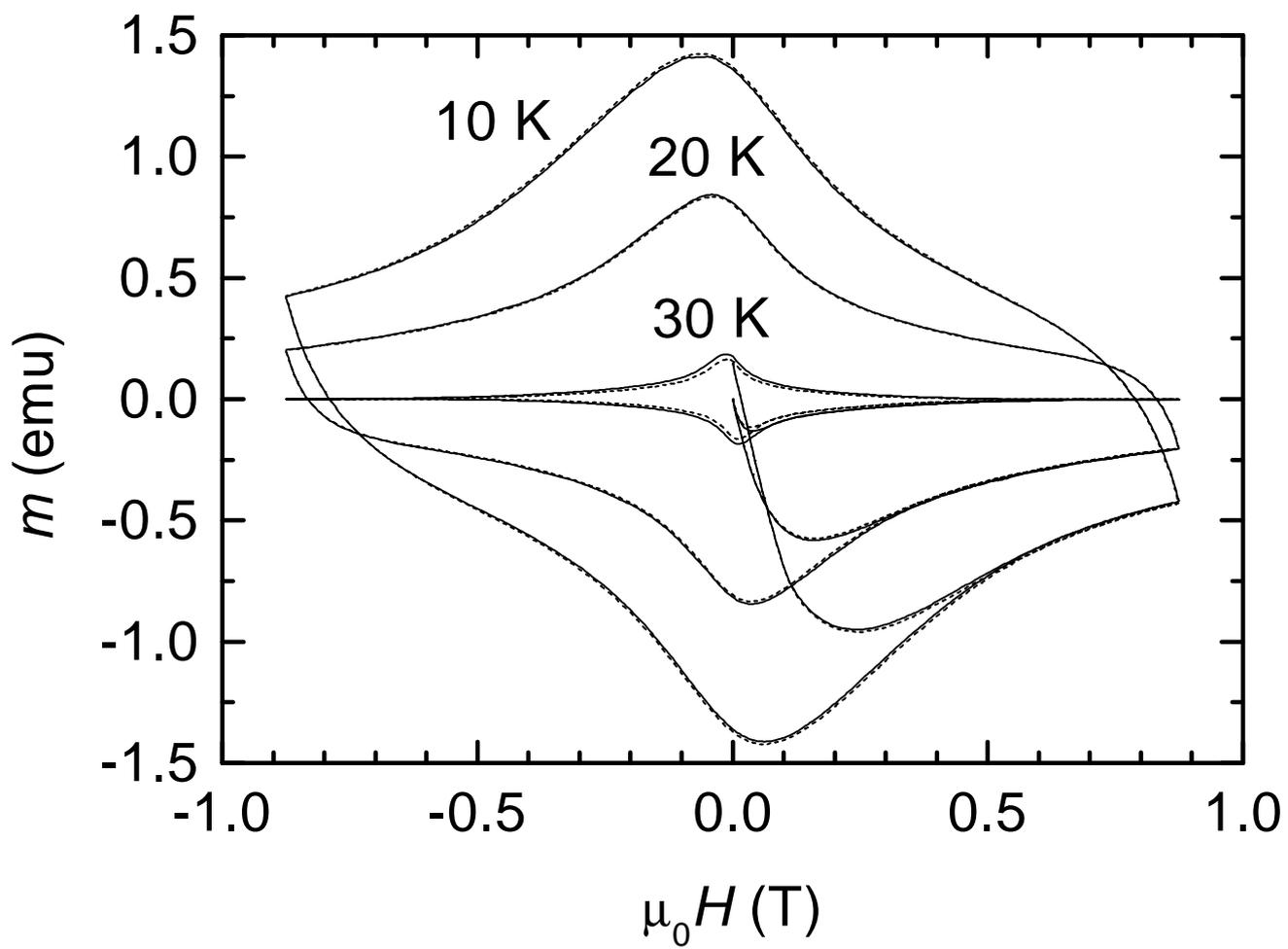
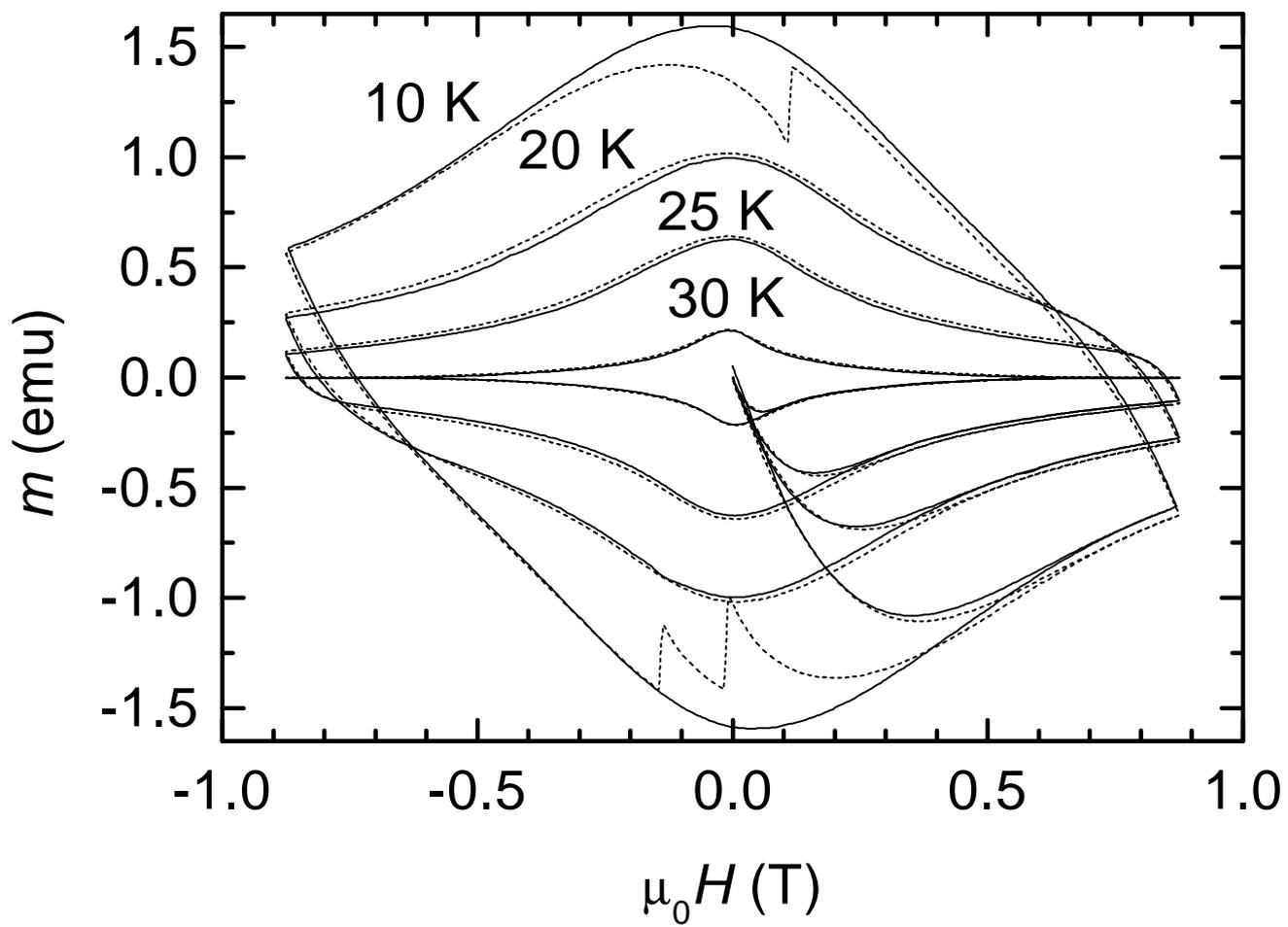
\end{document}